\documentclass[twocolumn,showpacs,preprintnumbers,amsmath,amssymb]{revtex4}
\usepackage{graphicx}% Include figure files
\usepackage{dcolumn}% Align table columns on decimal point
\usepackage{bm}% bold math
\usepackage{graphics}

\begin{document}

\preprint{APS/123-QED}

\title{Empirical signatures of shape phase transitions in nuclei with odd nucleon numbers}

\author{D. Bucurescu}
\email{bucurescu@tandem.nipne.ro} 
\author{N.V. Zamfir}  
\affiliation{Horia Hulubei National Institute of Physics and Nuclear Engineering, 
R-76900 Bucharest, Romania}
 
%e-mail: bucurescu@tandem.nipne.ro

\date{\today}
%% \footnotetext[1] {...}

\begin{abstract}
Nuclear level density at low excitation energies is proposed as an indicator of the first order
phase transitions in nuclei. The new signature, a maximum value of the level density 
at the critical point, 
appears to be sensitive to the detailed way the
phase transition takes place in different nuclear regions: it is consistent with phase coexistence
in the $N=90$ region, and with a rapid crossing of the two phases, 
without their coexistence/mixing at 
the critical point in the $N=60$ region, respectively. Candidates for critical point nuclei are
proposed for odd-mass and odd-odd nuclei, using correlations between relative excitation energies,
and their ratios, for structures (bands) based on unique-parity orbitals.  
\end{abstract}


\maketitle 
\section{Introduction}

   Quantum phase transitions (QPT) in nuclei appear at zero temperature, and represent rapid
changes in the equilibrium deformation of 
the ground state, induced by the variation of a nonthermal control parameter 
-- the number of nucleons, due to the competition of phases with different shapes. Such changes
in the shape of the ground state 
influence the evolution of various nuclear properties, and, in order to characterize this type of QPT
(called also shape phase transition -- SPT), it is important to identify observables that can play
the role of order parameters, allowing to observe how these critical phenomena 
take place. Because the natural control parameter in nuclei, the nucleon number, has only integer 
values, its discontinuous variation smoothes out the discontinuities at the phase transition point
(or critical point). 
Both experimental and theoretical studies propose various 
ways to assess the different signatures of the SPTs: the 
discontinuous behavior of nuclear properties that can be related to order parameters
and the characterization of the type of transition (first or second order), 
the coexistence of the two phases, and the possible realization of critical points in real nuclei.
There are several review papers which present the nuclear QPT domain
from phenomenological point of view, as well as based on theoretical nuclear structure 
models \cite{G-Cejnar,G-Casten1,G-Casten2,GT-Cejnar,GT-Iachello} such as the Interacting Boson Model 
\cite{IBM} or the geometric (or collective) model {\cite{BMM,GG}. 

Most of the theoretical and experimental studies 
concentrated until now
mainly on the even-even nuclei. The reason is that the evolution of many experimental 
observables of these nuclei can be followed over extended nuclear regions thus allowing to identify
and characterize the discontinuities that are typical of SPT. 
Also, in the even-even nuclei, in order to describe the phase transitional nuclei, 
Iachello introduced, in addition to the 
three IBM benchmarks of collective behavior (spherical nuclei -- or with U(5)  dynamical symmetry, 
$\gamma$-soft 
nuclei -- with O(6)  dynamical symmetry, and nuclei with axially symmetric deformation --
with SU(3)  dynamical symmetry) \cite{IBM},
 the concept 
of critical point symmetries. He proposed two such models, called E(5) for the second order phase transition 
(between spherical, U(5) nuclei and $\gamma$-soft, O(6) nuclei) \cite{E5}, and X(5) for the first
order phase transition (between spherical, U(5) nuclei and symmetrically deformed, SU(3) nuclei) \cite{X5}.   
In contrast, the odd-mass and odd-odd 
nuclei were much less studied until now, because of their more complicated structure. Experimentally,
it is difficult to find, for such nuclei, observables that can be followed over extended nuclear regions; 
theoretically, the dynamical situation created by extra-particle(s) added to the even-even core, is more
complex. Only during the last decade there has been a boost of studies (both theoretical and experimental)
devoted to these nuclei 
\cite{Iac-E5/4,Alonso1,Alonso-E5/12,E5/12,Boyuk,Jafar,Iac-Lev,Petrellis,Nomura1,Nomura2,Quan,Yu1,
Yu2,Yu3,Yu4,Yu5, Buc1,Buc2}.

In this work we further extend our previous empirical approach to SPT in odd-mass nuclei \cite{Buc1}, 
by focusing on experimental quantities that can be used as order parameters for all 
types of nuclei, both 
even-even and with odd numbers of nucleons.  
The nuclear level density at low excitation energies is proposed as a novel 
indicator for first order SPT. Other aspects of a SPT, such as phase coexistence and 
critical point nuclei are considered as well. 

\section{Empirical observables as effective order parameters} 

Because the equilibrium deformation of the ground state is not an observable, one recurses to the so-called 
effective order parameters, which are experimental observables sensitive to SPT occurrence. Reference  
\cite{Iac-Zam} addressed this 
problem by determining the order parameter (deformation) both classically, with a Landau-type of potential, 
and by quantum calculation within the IBM. This study emphasized that (i) the finiteness of the nuclei leads 
to a certain smoothing of the expected discontinuities but that this is not a dominating effect, and 
(ii) the advantage in the nuclear case 
is that properties of both ground state and excited states can be 
measured. One can then use as effective order parameters observables related both to the ground
state and to excited states, which may present distinct signatures for the characterization of the SPT.
Generally, the critical behavior of excited states may differ from that
of the ground state, and thus lead to a slightly different critical point in some cases. 

In the following we briefly review the effective order parameters used until now to identify signatures
of the SPTs. 
 
\subsection{Nuclear mass-related quantities} These may be considered the most basic ones 
because they are related to the 
ground state properties. Nuclear masses or binding energies comprise the sum of 
all binding effects in a nucleus (single-particle
and residual interactions). However, their values (in MeV) are very large and thus obscure 
the effects (of a few MeV)
produced by changing the nuclear deformation. It is therefore convenable to use nuclear mass-based 
quantities which are of a differential nature and consequently more sensitive to changes, 
such as different separation energies. 

One of the most useful 
such quantities is the {\it two-neutron separation energy} $S_{2n}$, proposed long ago in a paper
which studied the classical limit of IBM \cite{Dieperink}, and intensively used since then to identify 
and characterize SPTs (see \cite{G-Cejnar,G-Casten1,G-Casten2}, and, for a few recent references, 
\cite{Iac-Lev,Petrellis,Yu1,Yu2,Yu4,Buc1}). 
For a chain of isotopes, $S_{2n}$ shows an almost linear decrease with increasing $N$, with
discontinuities at the shell closures and at the critical points of first order SPTs. 
The change at the critical point shows up
as a flattening or even an increase of the curve,  which translates into a singularity/kink in the 
derivative of $S_{2n}$ with respect to $N$.  For the second order SPTs $S_{2n}$ shows a discontinuity 
only in its first derivative.
This behavior was discussed in many papers (see quoted references). 
The basic character of $S_{2n}$ as a direct signature of the SPT 
comes from its expression as $S_{2n}(Z,N) = B(Z,N) - B(Z,N-2)$, where $B$ is the binding 
energy (the energy of the ground state). 
It is therefore proportional to the first derivative of the nuclear binding
energy with respect to the order parameter $N$, and thus plays a role in the characterization of 
the QPT similar to that of the free energy and its
derivatives with respect to the order parameter in the classical case \cite{G-Cejnar,G-Casten1,Yu4}. 
The differential variation of 
$S_{2n}$ (therefore the second derivative of $B$)  peaks at the critical point of the 
$SU(3)$ to $U(5)$ transition, the peak being a $\delta$ function in the infinite-size limit
(infinite number of bosons), and showing a smoother behavior for realistic numbers of bosons \cite{G-Cejnar}.
 Other quantities expressing various mass differences, such as $\alpha$-particle 
decay energies $Q({\alpha})$, double-$\beta$ decay energies $Q(2\beta)$, as well as other decay energies or 
 $Q$-values, are also useful to characterize SPT in nuclei \cite{Yu1}. One should emphasize that 
such mass-related quantities as presented above are available for a large number of nuclei (as they are
derived from mass tables \cite{masstable}) and can be used to identify and characterize SPT in any kind
of nuclei (even-even or with odd $N$ and/or odd $Z$).

\subsection{Nuclear radii and related quantities} Nuclear charge radii are sensitive to shell/subshell closures, 
and also to changes in the ground 
state deformation (actually, they are directly related to the $\beta_2$ deformation parameter). 
Therefore, similar to the two-neutron separation energies, nuclear radii 
and their differential variation show discontinuities and kinks at shell closure and 
critical points, respectively \cite{Angeli,Angeli-Marinova}.
Measured nuclear charge radii are available for many nuclei of all types \cite{radii}. 
Quantities that were considered as signatures of
the SPT are the average squared charge radii $<$$r^2$$>$ and their differential variations, like 
the isotope shifts (differences between two adjacent isotopes with mass $A$ and $A+2$), and the
isomer shifts [differences for the states $0_1^+$ and $2_1^+$, or for the $0_1^+$ and $0_2^+$
states, respectively, in the same (even-even) nucleus]. The later are sensitive
to the type of the SPT (first or second order) \cite{Iac-Zam,McCutchan}.

\subsection{Quantities based on excitation spectra of nuclei} 

(1) Excitation energies, $E(2_1^+)$, $E(4_1^+)$, and their ratio $R_{4/2} = E(4_1^+)/E(2_1^+)$ (for 
the even-even nuclei) were largely used for a long time. Similarly, certain 
relative excitation energies and 
their ratios were also used in odd-mass nuclei \cite{Nomura1,Buc1,Buc2}. To distinguish the first- and second-order
SPT within the IBM, the ratio $E(6_1^+)/E(0_2^+)$ was proposed \cite{Bonatsos}.  

(2) Electromagnetic transition strengths. $E2$ transitions within the ground state band, 
especially $B(E2;2^+_1 \rightarrow 0^+_1)$ (which is related to the
deformation parameter $\beta_2$), and the ratio $B_{4/2} = B(E2;4^+_1 \rightarrow 2^+_1)/B(E2;2^+_1 
\rightarrow 0^+_1)$; or between different bands: $B(E2;2^+_2 \rightarrow 2^+_1)$, and 
$B(E2;2^+_2 \rightarrow 0^+_1)$ \cite{BE2-interband,Zhang-interband}. Another good
SPT indicator was found in 
the electric monopole, E(0) transition strength $\rho^2(E0:0^+_2 \rightarrow 0^+_1)$
\cite{Wood,PvB-E0,Wiederhold}.

(3) Two-neutron transfer reaction intensities (from the $0_1^+$ ground state of the mass $A$ target to the 
$0_1^+$ and $2_1^+$ states of the final $A-2$ or $A+2$ nuclei) \cite{2n-Fossion,2n-Yu}. 

(4) Global properties of the excited $0^+$ states in even-even nuclei \cite{0+}.
   
From the list presented above, one can see that most of the experimental quantities used until now
as signatures (or order parameters) of 
SPTs in nuclei refer to even-even nuclei. It is therefore of utmost 
importance to enlarge this list with other quantities that can be used for all types of nuclei.

\section{Nuclear level density: a novel indicator for first order shape phase transitions}

Previous results, both theoretical calculations and experimental findings, indicated that 
at the critical point of the transition between spherical and axially 
deformed shapes the excited nuclear states have a peculiar 
behavior. Thus, IBM calculations predicted that the spectrum of the low-lying 
excited $0^+$ states was 
maximally compressed at this point \cite{Cejnar-PRE}, this behavior persisting for higher spins
as well. Experimental studies of the two-neutron transfer reaction $(p,t)$, which is particularly 
suited to evidence $0^+$ states, showed that the nucleus $^{154}$Gd, known as a good example 
of X(5) critical point nucleus (for the transition from spherical to axially deformed nuclei) had indeed an
enhanced number of low-lying (up to 2.5 MeV) $0^+$ states, compared to other nuclei in the
same region \cite{0+,0+PRC}. These results have led to the expectation that, near the critical point,
there may be an enhancement of the density of levels at low excitation energies.

In order to investigate this possibility we examine now experimentally determined nuclear level densities. 
We refer to the results of Ref. \cite{vE-DB} for the nuclear level densities at relatively low 
excitation energies. In that work, the total level density $\rho(E)$ 
was determined for individual nuclei by fitting 
both the cumulative numbers of known low-lying levels and
the density of neutron resonances, with the 
back-shifted Fermi gas model formula (BSFG) \cite{BSFG-CT},
\begin{equation}  
\rho_{BSFG}(E)  = \frac{e^{2 \sqrt{a (E - E_1)}}}{12 \sqrt{2} \sigma 
a^{1/4} (E-E_1)^{5/4}}
\end{equation} 
with the free parameters $a$ and the energy shift $E_1$. 
The nuclei for which this procedure can be applied is limited to about 300, as they are of the
stable-nucleus-plus-one-neutron type. The fitting procedure is described in \cite{vE-DB}. 
The
result of the individual fits to all these nuclei (between $^{18}$F and $^{251}$Cf) are shown 
in the inset of Fig. 1.  The values of $a$ show an almost linear 
dependence on $A$, on which oscillations due to shell effects are superimposed
\cite{BSFG-CT,vE-DB,RIPL3}. 
 Note that for nuclei with comparable mass values, 
the larger the parameter $a$, the larger is the level density $\rho$,
therefore one can use $a$ as a measure of the level density.

\begin{figure}[t] 
\includegraphics[scale=0.45,angle=270]{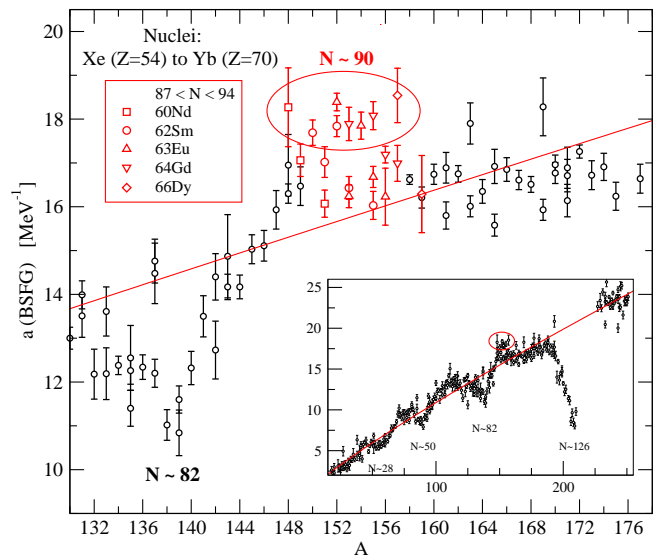}
\caption{\label{fig:epsart} Parameter $a$ of the BSFG model level density formula, eq. (1), as 
determined by individual fits to low-lying levels and neutron $s$-wave resonance spacings of about 
300 nuclei \cite{vE-DB}. The inset shows the values for all investigated nuclei, while the big 
graph displays the details for the nuclei with mass $\sim$ 150 and $N$ around 90. The red lines
represent the leading term of eq. (2), that is, $0.199A^{0.869}$.}
\end{figure}

By observing correlations between the parameters $a$ and $E_1$ and different experimental quantities, 
the following compact empirical formulas were proposed to describe these experimental values:
\begin{equation}
a ~=~ (0.199 + 0.0096 S') A^{0.869},
\end{equation}
\begin{equation}  
E_1  = -0.381 + 0.5Pa'
\end{equation} 
where $S' = S + 0.5 \cdot Pa'$, with the shell correction  $S = B_{LD} - B_{exp}$ 
being the difference 
between the binding energy $B$ calculated with a Weizs\" acker-type (liquid drop) formula \cite{Weiz} 
and the 
experimental value, and $Pa'$ is the so-called deuteron pairing
defined  as $Pa' = [2B(Z,N)-B(Z+1,N+1)-B(Z-1,N-1)]/2$, $B$-values as tabulated in 
 \cite{masstable}.
Formulas (2) and (3) can be used to predict level densities of nuclei for which these 
are not known or cannot be measured, by using only quantities from the mass tables.

 One remarks in the inset of Fig. 1 
a set of nuclei with mass $A\approx 150$ which significantly deviate from the average trajectory
of the data. The big graph of Fig. 1 presents in detail the region with these nuclei. 
 One  observes that the isotopes of
Nd to Dy with $N$ around 90 have larger level densities than other nuclei from this mass region, their 
$a$ values being about 18 MeV$^{-1}$, compared to about 16.5 MeV$^{-1}$ of the average behavior. 
These are exactly the nuclei near/at the $N=90$ critical point of the well known transition from spherical
to axially deformed nuclei in this mass region. 
Therefore, another signature of the first order SPT from this region is an increase of the level
density at the critical point.
 The examination of the experimental level densities at the neutron resonance energy brings 
further support to this conclusion. Figure 2 shows the experimental values of the 
density of the $s$-wave neutron resonances $\rho_0$, as determined from the average 
level spacings of these resonances \cite{RIPL3}.
The same set of nuclei, around 
the critical point at $N=90$, shows almost the largest $\rho_0$ values 
observed in the $\approx 300$ nuclei, some of them 
comparable  to values observed only in some transactinide nuclei. 
\onecolumngrid

\begin{figure}[t] 
\includegraphics[scale=0.6,angle=270]{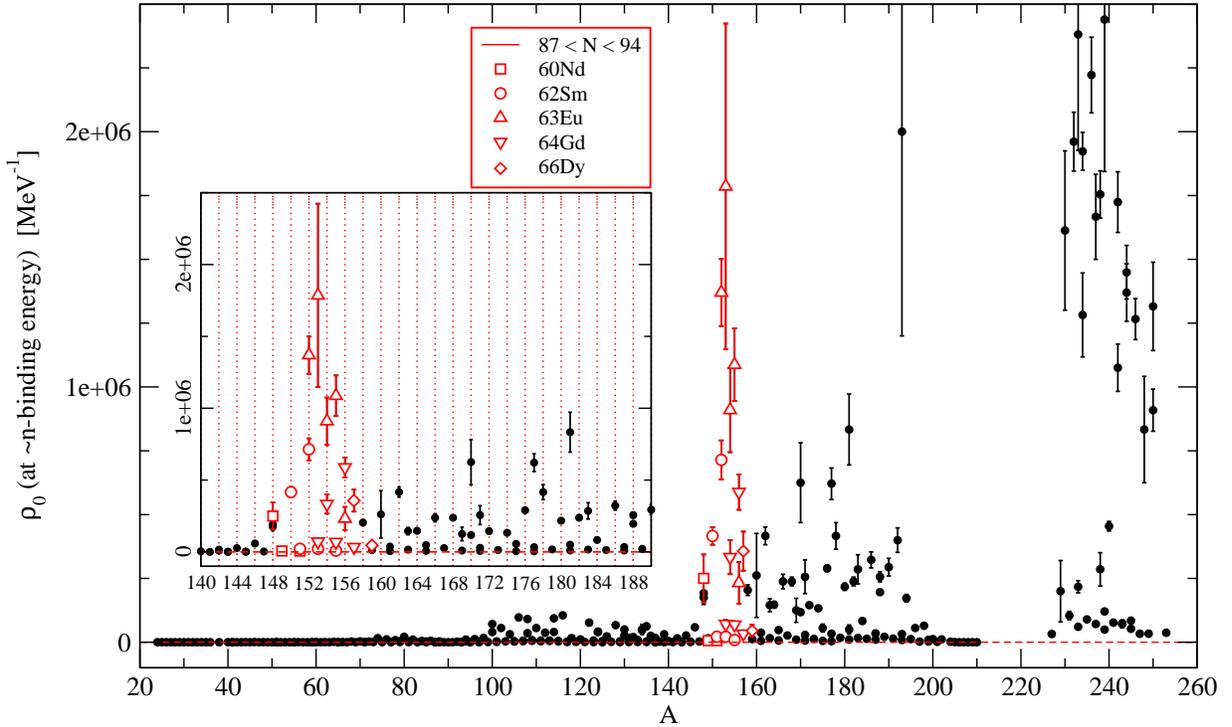}
\caption{\label{fig:epswide} Level density $\rho_0$ of the $s$-wave neutron
resonances for all nuclei from Fig. 1
(determined as the inverse of the mean resonance spacings from the RIPL-3 database \cite{RIPL3}). 
The inset shows in detail the nuclei 
in the region of ths shape phase transition at $N=90$.}
\end{figure}
\twocolumngrid

\twocolumngrid
 One should remark here that the set of about 300 nuclei represented in both 
 Figs. 1 and 2, 
for which experimental $a$ and $\rho_0$ values are known, is rather restricted, as it comprises only
nuclei that can be obtained from stable targets plus one neutron. The SPT region with the 
peak around the 
$N = 90$ critical point shows up so well because it contains, fortunately, 
many such nuclei, thus allowing to see a 
meaningful correlation between the SPT and the level density 
both at low excitation energies and at the neutron resonance energy. 
However, many key nuclei from this region,  close to the 
critical point,  cannot appear in these plots (such as, e.g.,  $^{150}$Nd), and this is 
generally the case for other nuclear regions where first order SPTs occur. 

 In order to see how this first order SPT occurs as a function of the neutron number $N$ 
 (the control parameter)
in the level densities of all nuclei from the mass $\approx$150 region we rely on 
formula (2) for the parameter $a$, which describes reasonably well the set of the 300 
experimentally investigated
nuclei \cite{vE-DB}. Figure 3 shows how $a$ evolves as a function of  $N$ for 
all types of nuclei with $Z$ between 56 (Ba) and 70 (Yb). The curves passing through small symbols 
are the values predicted by Eq. (2). One can see that they describe reasonably well the few 
existing experimental data from this region 
(the big symbols, representing the values obtained by individual fits to those 
nuclei \cite{vE-DB}), which suggest themselves the existence of maxima around 
$N{\approx}90$. 
For all four types of nuclei (with both even and odd numbers of nucleons) the curves of the isotopic 
chains between $Z=60$ (Nd) and $Z=67$ (Ho) present a maximum for $a$ (maximum level density) 
around the critical value $N=90$. For isotopes with larger $Z$
this maximum gradually shifts to larger $N$ and diminishes, while for smaller $Z$ it does not exist.
It is thus seen that the main correction in Eq. (2) to the almost linear dependence on $A$, which is 
due to the "shell correction" $S$, contains information not only on the shell closure but on the
occurrence of the shape phase transition as well -- which 
is not surprising, as it is determined by the nuclear masses. 

As Figs. 1 to 3 show, the nuclear level density at low energies, represented by the main 
parameter $a$ of its BSFG model description is a good signature for the SPT at $N=90$,
the nuclei near the critical point having the largest level densities. This confirms the
predictions of the IBM calculations for the maximization of $0^+$ states density at the 
U$(5)$ to SU$(3)$ transition point \cite{Cejnar-PRE}, and corroborates 
the experimental findings for the numbers of $0^+$ states in nuclei from the same region
\cite{0+,0+PRC}. 
The meaning of this level density enhancement 
at relatively low energies may be related to the special potential well
of the nuclei near the critical point  \cite{G-Casten1,G-Casten2}. 
It is expected that in 
such nuclei the potential well is broadened (in the deformation space) due to the nearly 
degenerated
spherical and deformed coexisting minima, with a small barrier between them. 
As states can
exist in either minimum, such a potential well can accommodate more states, their large density
and the small barrier favoring the state mixing. Therefore,  
 the enhancement of the level density, clearly shown in the presented data for the mass 
 ${\approx}150$ region,
 is consistent with the {\it phase coexistence}, another 
phenomenon expected for a first order SPT.
\onecolumngrid

\begin{figure}[h]
\vspace{13 mm}
\includegraphics[scale=0.6,angle=270]{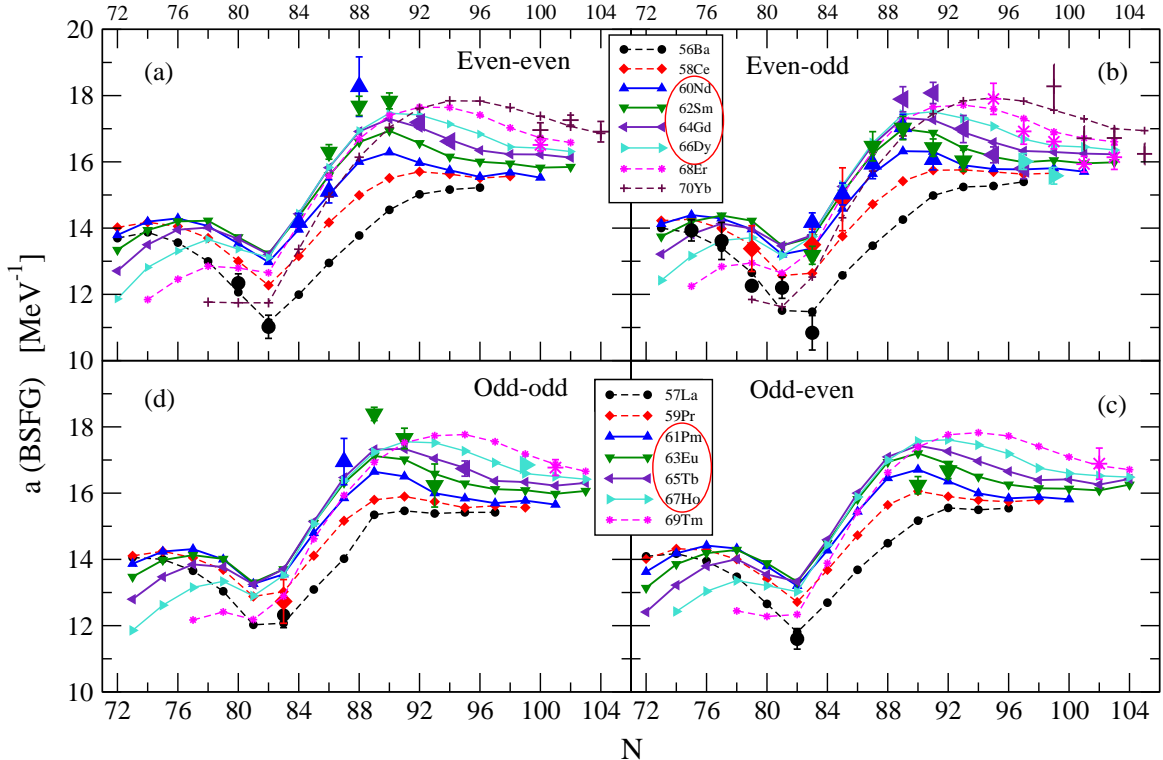}
\caption{\label{fig:epswide} Level density parameter $a$  for nuclei in the
region of the SPT at mass $\approx$150. Small symbols joined by curves are values calculated with
formula (2). Big symbols are experimental values determined by 
individual fits with formula (1) \cite{vE-DB}.}
\end{figure}
\twocolumngrid

%\vspace
\section{Comparison between the $N=90$ and $N=60$ SPT regions}

It is interesting to apply the present approach based on the level density 
to other well established nuclear regions showing a 
first order SPT. One of these, well known for its very rapid transition,  
is the $A\approx 100$ (Sr - Zr) region, 
with the critical point at $N = 60$. 
Before looking at level densities, we first make a comparative analysis of the 
the $A\approx 150$ and 
$A\approx 100$ regions, using
general order parameters that can be applied to all types of nuclei, that is, the quantities 
$S_{2n}$ and $<$$r^2$$>$ discussed in Sec. I. Actually, we have used their differential variation
(difference between two successive isotopes) which acts as a "magnifying glass" in highlighting the 
discontinuities associated with the SPT.

Figure 4 displays the differential variation  $dS_{2n}$ of the two-neutron separation energy, and    
$d$$<$$r^2$$>$ of the mean square charge radius, respectively,
for the nuclei in the region of the STP at $N=90$, separated
for the four types of nuclei according to their nucleon numbers.  $S_{2n}$  can be 
examined for both even-even
nuclei and  nuclei with odd numbers of nucleons, as shown in some recent papers \cite{Petrellis,Yu1,Yu2,Yu4,
Buc1}. Its behavior in this mass region was considered in detail in Refs.  \cite{Petrellis,Yu1,Yu2,Yu4}, 
by examining its deformation-dependent part, as well as the odd-even effects. It was found that the signal 
for the emerging SPT is enhanced by the extra-single particle(s) in comparison with that from the adjacent 
even-even nuclei, and that pairing plays the major role in driving the phase transition. The enhancement 
of the signal for the SPT can be also seen in graphs (a) to (d) of Fig. 4, 
the amplitude of the kink at $N \approx 90$ being larger for the odd 
nuclei than for the even-even ones, with the largest effect in the odd-odd nuclei, in agreement with the
findings of Ref. \cite{Yu1}. A recent study of Sm and Eu nuclei, with a core-quasiparticle coupling 
Hamiltonian based on the energy dependence functionals, explained the enhancement of the SPT in the odd-mass 
nuclei by a shape polarization effect of the unpaired proton, which, before the critical neutron 
number 90 couples to  Sm cores, and starting from $N=90$ couples predominantly to Gd cores which have larger 
quadrupole deformation and smaller pairing \cite{Quan}. 

\begin{figure} 
\includegraphics[scale=0.47,angle=0]{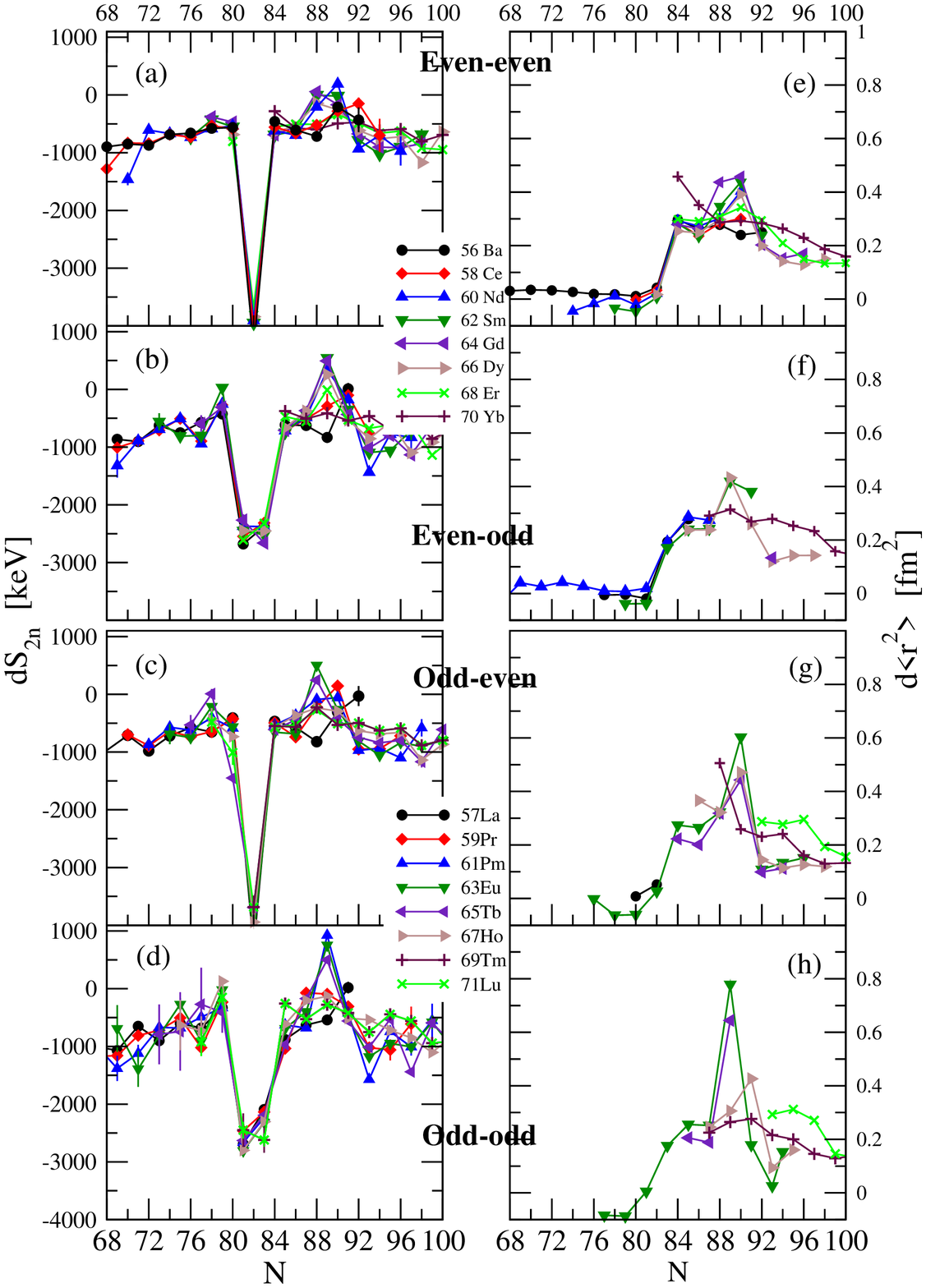}
\caption{\label{fig:epsart} Differential variations of the two-neutron separation energy $S_{2n}$ 
(data from \cite{masstable}) and of the 
mean square charge radius $<r^2>$, or isotopic shifts (data from \cite{radii}),
 for isotopic chains in the $A\approx 150$ region,  around $N = 90$. 
The definitions used are 
$dS_{2n}(Z,A)$=$S_{2n}(Z,A+2)$-$S_{2n}(Z,A)$ and 
$d<r^2(Z,A)>$=$<r^2(Z,A)>-<r^2(Z,A-2)>$, respectively, and
were chosen such as to display a correct behavior at the magic number $N=82$.}
\end{figure}

Graphs (e) to (h) in Fig. 4 show the differential variation of the mean square charge radius for 
the same nuclei from the left column of the figure, whenever these quantities are
available \cite{radii}.
The aspect of these graphs in the region $N \approx 90$ is similar to that of the corresponding graphs from the
left side, also showing the enhancement of the signal in the nuclei with odd number of protons. 
Actually, 
a similarity between the graphs of various differential observables (these two and 
those of some other spectroscopic
observables available only  for even-even nuclei) as a function of $N$, 
was remarked in Ref. \cite{Cakirli}.

Figure 5 is the analog of Fig. 4, for the region with mass $A \approx 100$, with its fastest 
known shape phase transition at $N=60$. It looks rather similar to Fig. 4, except for the smaller number of 
isotope chains which show the SPT in the graphs (a) - (d) of $dS_{2n}$.
There is a clear signature in 
even-even nuclei [graph (a)] only for Sr and Zr (also well known from other spectroscopic observables).
 In the even-odd nuclei [graph (b)] the signature for the first order SPT is enhanced  for Sr and Zr 
and there is also a weaker signature for the Mo isotopes. 
In the nuclei with odd proton [graphs (c) and (d)]
there is indication of SPT in Rb, Y, and Nb nuclei, with an enhancement of the signature 
in the odd-odd nuclei.
Graphs (e) to (d) for the charge radii 
show a strong SPT signature, unfortunately the only chains with available data
being Sr and Zr [graphs (e) and (f)], and Rb and Y [graphs (g) and (h)].

 \begin{figure} 
\includegraphics[scale=0.47,angle=0]{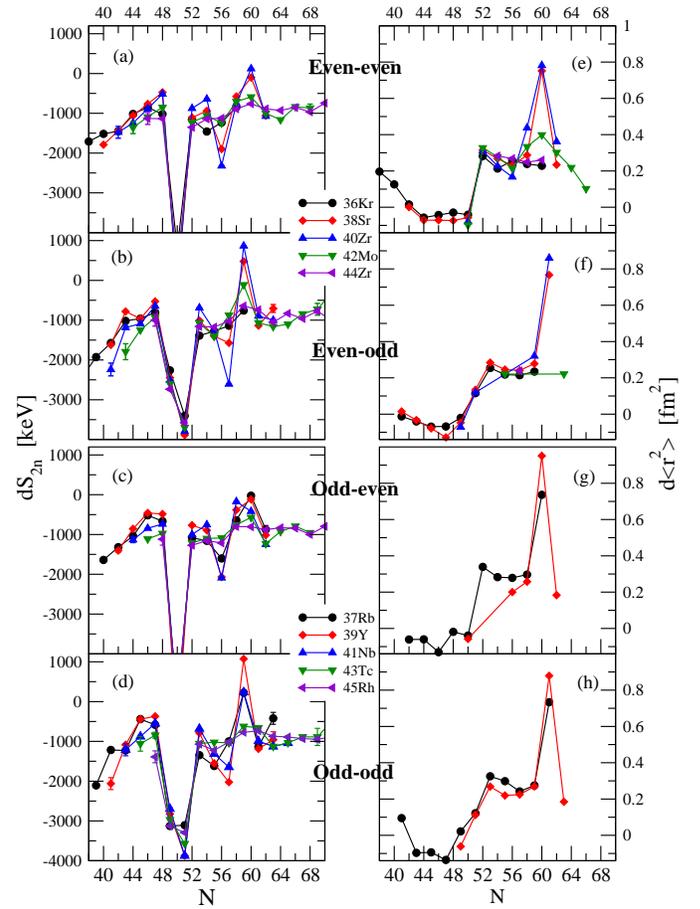}
\caption{\label{fig:epsart} Same as Fig. 4, for the $A\approx 100$ region, around $N =60$.}
\end{figure}

Figures 4 and 5 show a rather close similarity between 
the two regions, with clear signatures of  first order SPT appearing at 
$N=90$ and $N=60$, respectively. 
We examine now how the SPT in the Sr -- Zr region appears in the evolution
of the level density. Figure 6 is similar to Fig. 3, it displays the level density parameter $a$ for the 
isotopic chains from Kr to Rh. 
Unfortunately, there are no experimentally determined 
level densities for the rather neutron-rich nuclei around the critical point $N=60$.
There are only
several nuclei with experimental points (big filled symbols), all below $N=60$, the trend of their 
evolution with $N$ being reasonably well reproduced by Eq. (2). We rely therefore 
on formula (2) to examine how the level density evolves in this region. 
It is  likely that this formula predicts with reasonable accuracy
the level density parameter $a$ at least for the cases when experimental masses are used 
(in Fig. 6 the
points calculated from extrapolated masses -- from Ref. \cite{masstable}, are distinguished 
by open small symbols and dashed curves).
The shape phase transition (kinks in Fig. 5) is indicated by a discontinuity in the evolution 
of $a$, shown by  the
 change of slope at $N=60$, after which the values remain almost constant (Fig. 6). 
\onecolumngrid

\begin{figure}[h] 
\vspace{13mm}
\includegraphics[scale=0.6,angle=270]{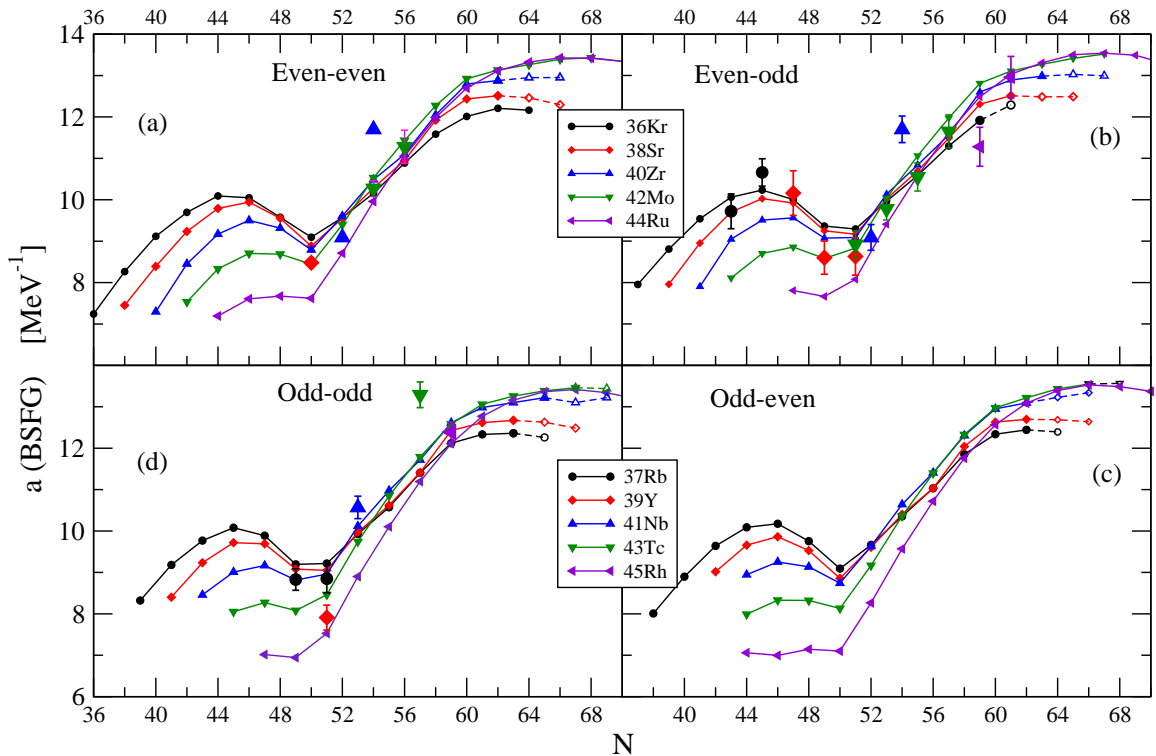}
\caption{\label{fig:widefigure} Same as Fig. 3, for the $A\approx 100$ region. Small
empty symbols joined by dashed lines indicate values calculated with Eq. (2) 
from extrapolated mass table values \cite{masstable}.}
\end{figure}
\twocolumngrid

This behavior is different 
from that of the nuclei from the Sm region (Fig. 3). 
The lack of a maximum of the level density at the critical point $N=60$ in Fig. 6 may indicate
 that in the Sr -- Zr region there is no phase coexistence at the
critical point. This situation corroborates the results of recent large scale 
Monte Carlo shell model 
calculations which
describe very well the characteristics of nuclei in this region, namely, $^{96}$Zr \cite{96Zr} 
and the 
SPT at $N=60$ \cite{100Zr}. In all nuclei from this region, from $^{96}$Zr to $^{110}$Zr, 
these calculations
predict coexistence of various shapes: spherical, prolate, oblate, and triaxial, 
the mechanism 
describing quantitatively the low-energy structure of these nuclei being 
the type II shell evolution
\cite{,typeII,96Zr,100Zr}. The SPT  in the Sr -- Zr region is a rather special one, the transition
from nuclei with spherical ground state to nuclei with axially deformed ground state 
being a rather abrupt one, 
the two competing configurations crossing each other without significant interaction between 
them. The same situation appears in the similar transition from $^{96}$Sr and $^{98}$Sr, where
Coulomb excitation studies showed that the prolate (g.s.) and spherical (excited) 
coexisting configurations of $^{98}$Sr have a very weak mixing \cite{96-98Sr}.
In the Sm region
this transition is comparatively more gradual, and takes place through a critical point where the 
two phases coexist and mix with each other. A discussion of the 
shape coexistence and 
phase coexistence phenomena can be found in \cite{G-Cejnar,phase-shape}. 
We conclude this section by stating that the analysis of these two nuclear
regions shows that shape phase transition and phase mixing in nuclei 
can appear in different forms 
and with different characteristics, as also remarked in \cite{BE2-interband}. 

\section{Critical point nuclei}    

Besides the abrupt change in the order parameter (ground state equilibrium deformation) 
and the phase coexistence phenomenon, a third 
feature of the SPT in nuclei is the possible experimental occurrence of 
 critical point nuclei. Because the number 
of nucleons is an integer, the nuclear properties change discretely, and in principle 
it may happen that real nuclei do not lie exactly at the critical point. 
The identification of nuclei with critical point features in the even-even case has mainly been based on 
comparison of spectroscopic observables with  the parameter-free predictions of the critical point
symmetries E(5) \cite{E5} and X(5) \cite{X5}. 
Thus, for the first order SPT a number of nuclei with X(5) 
properties were proposed in the rare earths region, in particular the 
$N=90$ nuclei $^{150}$Nd \cite{150Nd}, $^{152}$Sm \cite{152Sm}, $^{154}$Gd \cite{154Gd},$^{156}$Dy \cite{156Dy},
as well as other possible candidates in heavier 
rare earths \cite{Clark,166Hf}.  IBM calculations also allowed localization of the critical
phase transition points (see, e.g., Ref. \cite{Ramos-IBM}).
For the odd-mass nuclei, there are also theoretical developments
of critical point symmetries. Thus, Iachello introduced the critical point Bose-Fermi symmetry 
E(5/4) for the case of a $j=3/2$ particle coupled to an E(5) core \cite{Iac-E5/4}, subsequently
extended into the E(5/12)  model \cite{Alonso-E5/12,E5/12} by considering a multi-$j$ case: coupling
of a particle in $j = 1/2, 3/2, 5/2$ orbits. 
The X($5/(2j+1)$) model was proposed for cases 
in which a particle in a $j$ orbit is coupled to an X(5) core \cite{Yu3}. 
As empirical studies of different order 
parameters show (see also previous sections of this work), the first order SPTs 
in odd-mass nuclei are strongly correlated to those 
taking place in the adjacent even-even nuclei. 
The role
of the unpaired particle on the SPT was also studied within the  Interacting Boson-Fermion model
(IBFM) \cite{IBFM} calculations \cite{Alonso1,Boyuk,Iac-Lev,Petrellis}, where the theoretical 
results are compared to experimental evidence on the occurrence of phase transitions 
in Pm, Eu, and Tb proton-odd nuclei. The IBFM study of a particle in the $j=1/2,3/2$ and 5/2 orbitals 
\cite{Alonso1}
showed that the position of the critical point in the odd nucleus is shifted with respect to that
in the even-even core, with a magnitude proportional to $1/N$ ($N$ the number of active bosons). 
Newer developments refer to  
a microscopic framework based on the nuclear
energy density functional theory. In Ref. \cite{Nomura1}, the energy density functionals method and 
a fermion-boson coupling scheme were used,  
 and calculated spectroscopic observables indicate sharp 
irregularities at mass 151 for the Eu isotopes, and at mass 153 for the odd Sm isotopes. This type 
of study was recently refined, by using a core-quasiparticle-coupling Hamiltonian based on energy
density functionals \cite{Quan}. 

Compared to the even-even case, approaches to describe the structure of odd-mass nuclei
in a global way are only at their beginning. 
Besides studies  highlighting the role
of the unpaired particle on the SPT (usually manifested in an enhancement of the transition), 
 there is also interest, at present, 
in empirically finding  critical point nuclei with odd $N$ and/or odd $Z$. 
Thus, in Ref. \cite{Yu2}, it is proposed to identify critical point odd-mass nuclei by looking 
for the phase coexistence phenomenon, which should take place around the critical point of a first
order SPT. In that work the experimentally 
observed low-lying structure in the odd Sm nuclei was examined.
The several known low-lying bands known in these nuclei show that in $^{151}$Sm  
there are coexisting rotational and vibrational bands (phases), 
whereas the structure  of $^{153}$Sm is 
dominated by rotational structures. Therefore, it was proposed that, because $^{151}$Sm shows more 
clearly shape coexistence features, it is closer to the critical point \cite{Yu2}.    

For the odd-mass nuclei, a method based on correlations between excitation energies 
  has been rather useful to identify 
critical point nuclei. This kind of analysis was first proposed by Ref. 
\cite{AHV}, and represents an empirical  way to circumvent the integer nucleon number 
problem \cite{finiteN}. It consists in using as a control parameter an empirical quantity, such as 
$E(2^+_1)$ in even-even nuclei, which, for many nuclei in a region, presents a 
nearly continuous distribution. Other quantities [like $E(4^+_1)$, $S_{2n}$], represented as a function
 of $E(2_1^+)$,  follow simple, compact trajectories, for large regions of nuclei, 
with distinct anomalies, characteristic of a phase transition,  
at $E(2_1^+$) values similar to that of $^{152}$Sm ($\approx$150 keV) where phase coexistence
was suggested  \cite{AHV}. 
The same is true for other differential observables \cite{Wolf}. A similar procedure was subsequently 
proposed for odd-mass nuclei \cite{Buc1}.  In order to be able to cover large nuclear regions, 
the structures (bands) determined by unique parity orbitals (UPO) 
were considered because they practically do not mix with 
other orbitals and thus lead to nearly identical effects for any UPO. 
Excitation energies relative to that of the state of 
spin $j$ (the spin of the UPO) were defined within the favoured quasiband  $E(j+2),~E(j+4)$, 
with $E(I) = E^*(I)-E^*(j)$, and also the ratio $R_{j+4/j+2} = E(j+4)/E(j+2)$, similar to 
$E(2_1^+)$, $E(4_1^+)$, and $R_{4/2}$ for the even-even nuclei.  $E(j+4)$ 
displays compact trajectories when represented as a function of $E(j+2)$ 
(the effective control parameter),
for the different investigated UPOs \cite{Buc1}. In particular, for the nuclei of mass 
$\approx$ 150, where 
there are rich data for the ${\nu}i_{13/2}$ UPO, this correlation has 
a "turning point" 
for nuclei near $N\approx90$ at a  minimum energy $E_c(j+2)\approx 200$ keV, 
which is correlated with the critical point of
their even-even cores, with $E_c(2^+_1)\approx 150$ keV \cite{Buc1}. 
In the odd-mass nuclei
this kind of correlation allows a rather direct identification of possible critical point nuclei, 
based on their proximity to the critical (turning) point $E_c(j+2)$. Figure 7 shows the
correlation between $R_{j+4/j+2}$ and $E(j+2)$ of ${\nu}i_{13/2}$ structures 
in the even-odd isotopes from Sm to Os. 
Although less compact than the correlation of $E(j+4)$ versus $E(j+2)$ (Fig. 6 of \cite{Buc1}), 
it allows a better visualisation of each isotopic chain. 
In Fig. 7, for each
isotopic chain, with increasing mass,  $R_{j+4/j+2}$ (and $E(j+4)$ \cite{Buc1}) first increases
as $E(j+2)$ decreases, this trend reversing at the critical (turning) point at $E_c(j+2)\approx
 200$ keV.
One should remark that the maximum of compression of the favoured band at the critical point
corroborates well the maximum level density at the same point discussed in Sec. II.
Based on the individual isotopic curves in Fig. 7, one can propose critical point 
nuclei as those which
are closest to the point where the curve reverses its trend (turns back in energy). In some cases, 
like $^{155}$Gd or $^{157}$Dy,  the experimental point coincides rather well with the
turning point of an empirical continuous curve drawn through the data points, 
therefore these nuclei 
are good candidates for critical points. In other cases, the situation may be different; for
example, for the Sm isotopes, $^{153}$Sm has the lowest $E(j+2)$ energy, 
but a continuous curve drawn
through the data points may suggest that the turning point occurs somewhere in between $^{153}$Sm 
and $^{151}$Sm.
Another feature of the critical point nuclei may be the approximate degeneracy 
in energy of the favoured and unfavoured UPO bands 
[$E(j+1) \approx E(j+2)$, $E(j+3) \approx E(j+4)$, ...] \cite{Buc1,Buc2}.
  
\begin{figure}[t] 
\includegraphics[scale=0.35,angle=270]{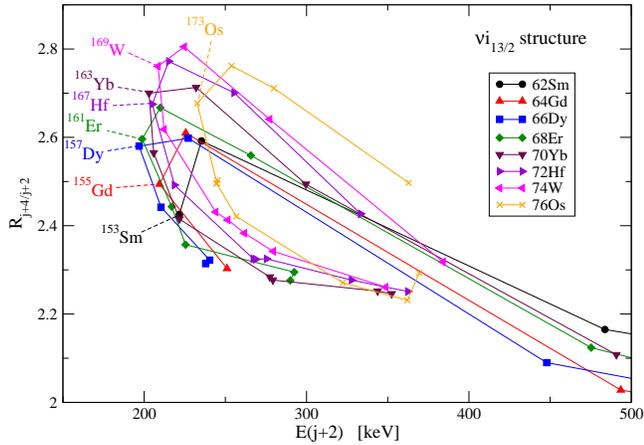}
\caption{\label{fig:epsart}Correlation between the energy ratios $R_{j+4/j+2}=E(j+4)/E(j+2)$ and the 
relative energies $E(j+2)$ for the favoured band of the ${\nu}i_{13/2}$ structures in the odd-mass 
nuclei of the isotopic chains
from Sm to Os. The nuclei closest to the turning points of the curves are explicitly indicated.  }
\end{figure}

Figures 4 and 5 show a very clear SPT signature for the odd-odd nuclei (Pm, Eu, and Tb in the 
mass-150 region, and Rb and Y in the mass-100 region, respectively). The complicated 
structure of these nuclei makes it difficult to follow observables related to excitation energies.
In the case of Tb ($Z=65$) one could, nevertheless, identify and follow the evolution of 
the same structure, namely  a
band with the $({\pi}h_{11/2},{\nu}i_{13/2})$ structure \cite{ENSDF}, 
for three nuclei around $N=89$ where
one can see the strong SPT signature in Fig. 4. 
Figure 8(a) shows a correlation between excitation energies within this band, 
relative to the state of spin 12$\hbar$ (which is the sum of the spins of the two orbitals), similar 
to the case of UPO states in the odd-mass nuclei. One can see that the evolution of the three 
nuclei shows a "turning point" at $^{154}$Tb ($N=89$), which may be proposed as a candidate for
critical point nucleus.
Figure 8(b) shows that this band is maximally 
compressed at $N=89$, similar to the odd-mass nuclei, and 
in agreement with expectations based on the level density argument. 
One should also remark that the same argument, of largest level density at the 
critical point nuclei, seems to be confirmed also for the $N=89$ nucleus $^{150}$Pm (Fig. 4), for 
which a large density of low-lying excited levels was observed through the $(d,\alpha)$ reaction, 
unlike for other nuclei in the same mass region (cf. Fig. 1 of \cite{150Pm}).
As Figs. 4 and 5 show, further experimental and theoretical studies of 
signatures of the first order SPT in odd-odd nuclei in the 
mass regions $\approx 150$ and $\approx 100$, should concentrate on the best candidates for
critical point nuclei $^{150}$Pm, $^{152}$Eu, $^{154}$Tb, 
and $^{96}$Rb, $^{98}$Y, $^{100}$Nb, respectively.

\begin{figure}[t] 
\includegraphics[scale=0.48,angle=0]{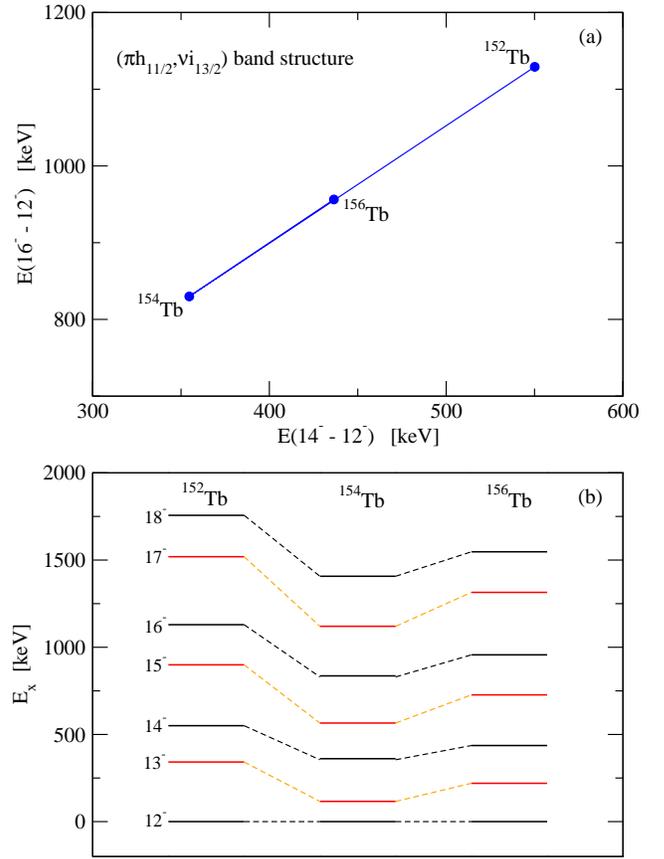}
\caption{\label{fig:epsart}(a) Correlation between relative excitation energies of the favoured 
sequence of the $({\pi}h_{11/2},{\nu}i_{13/2})$ structure in the odd-odd Tb ($Z = 65$) isotopes 
around $N=89$ (data from
\cite{ENSDF}; (b) The $({\pi}h_{11/2},{\nu}i_{13/2})$ band in the three Tb isotopes from graph (a). }
\end{figure}

\section{Conclusions}

In this work it has been shown that the nuclear level density at low 
excitation energy constitutes a good 
indicator for the first order shape phase transitions in nuclei. The level density displays a 
maximum value at the critical point, and this was tested in detail for the known SPT at $N=90$.
This maximum value also discloses the phase coexistence phenomenon at the critical point.  
It is gratifying that this novel indicator can be 
studied for all kinds of nuclei, even-even, odd-mass, and odd-odd.   

A comparison has been presented between two well known first order SPTs, those at $N=90$ and $N=60$,
by using effective order parameters that are applicable to all types of nuclei: the differential 
variation of the two-neutron separation energy $S_{2n}$, of the mean square radius $<r^2>$, and the 
level density (as represented by the $a$ parameter of its BSFG description). While the first two
such quantities are very similar in the two nuclear regions, just indicating a rapid change of  
properties, the one based on the level density appears as more shaded, as it 
indicates some differences related to the way the two SPTs take place: while the transition at $N=90$
is more gradual and indicates phase coexistence and mixing, 
the one at $N=60$ is consistent with a very rapid 
crossing of the two phases, without coexistence and mixing between them.

Finally, we discuss the problem of assessing critical point nuclei in the case of the odd-mass and
odd-odd nuclei. For the odd-mass nuclei, the method is based on the correlations between 
excitation energies, and their ratios, of the unique parity orbital structures (bands) \cite{Buc1,Buc2}.
It is shown, by examining the ratio $R_{j+4/j+2} = E(j+4)/E(j+2)$ versus $E(j+2)$
(with $j$ -- the unique parity orbital spin), that candidates for the critical point nuclei can be 
proposed by looking at their proximity to the critical energy ($E_c \approx 200$ keV in the case of 
nuclei with ${\nu}i_{13/2}$ bands from the $N=90$ region), which appears as a turning point in 
these graphs. 
The pattern of these correlations can be used to tell whether the critical point is very close to
a certain real nucleus, or it falls between two real nuclei. An example of 
such correlations is also given for a restricted number of odd-odd nuclei from the same region. 
Both the odd-mass and odd-odd nuclei display a maximum compression of the band structures near 
the critical point of a first order SPT, in 
agreement with the other criterion, of maximum level density. Some candidates of critical
odd-odd nuclei are proposed on the basis of the employed empirical investigations.       

\acknowledgments 
Partial funding by the Romanian project IFA - CERN-RO/ISOLDE is acknowledged.

\end{document}